# A TWO-QUBIT CELL ON THE BASIS OF BORON NITRIDE NANO-TUBES FOR THE QUANTUM COMPUTER

*M.B.Belonenko\*), N.G.Lebedev, I.V.Zaporotskova*

*Volgograd State University,*

*2-ya Prodolnaya Street, 30, Volgograd, 400062, Russia,*

*E-mail: nikolay.lebedev@volsu.ru*

*\*) Volgograd State Architectural-Building Academy,*

*Academical Street, 1, Volgograd, 400074, Russia*

1. The interest to quantum computers allowing at the expense of quantum parallelism significant to speed up of calculation, has appreciably increased after the basic papers [1-3], in which the effective quantum algorithms are described. Among various physical realizations of the quantum computer it is interesting model on the basis of a quantum crystal representing an one-dimensional chain of ions (for example, $Ca^+$, $Hg^+$), which have kept along an axis of a chain by locking potential, and in a cross axis of a chain directions by combinations of constant and variable electrical fields similar used in traps of Paul [4]. The experimentally received quantum crystals from 11 ions now are known. At the same time, it is well known, that it is necessary to make operation NOT above one qubit, and operation CONTROLLED NOT above two qubits for construction of the quantum computer, or, in other words, there should be an interaction between physical systems realizing of qubits. In discussed above model this role is played by indirect interaction of ions through the lowest oscillatory mode of a quantum crystal [5]. Also there are main difficulties on the ways of experimental realization of the full value computer (i.e. containing order 1000 qubits). On our sight, the main difficulties are: the high requirements to accuracy of the task of constant and variable fields in traps such as Paul's traps; the cross instability of an one-dimensional quantum crystal (i.e. propensity of a crystal to formation of "zig-zag"-type configurations); the requirement of deep laser cooling of ions for the maintenance of interaction through the lowest oscillatory mode. At the same time possibility of the individual manipulation to the separate qubit, the ability is easy to ensure the initialization of the quantum computer make doubtless advantage of the given model. In connection with above-stated also has taken place the stimulus of the given article directed on the solution of mentioned difficulties.

2. Recently one became widely to investigate boron nitride nanotubes (tubulenes), representing a hexagonal lattice of B and N atoms turned in the cylinder [6]. The geometrical structure of boron nitride nanotubes is characterized by two parameters (m, n), where m is responsible for the chiral of tube and parameter n determines number of hexagons on perimeter of tube. The ability of nanotubes to draw in itself atoms of various elements (the capillarity) [7] is now well known. So we'll consider updating quantum crystal consisting the ions (for example, $Be^+$, $Mg^+$ and etc.) are located inside boron nitride nanotubes by means of the capillarity effect and by application of the electrical field along nanotube axis. Locking potential for ions can be created just as by the application of external electrical fields, so and doping of nanotube by the appropriate ions. In this case constant and variable fields for keeping of ions perpendicular to nanotube axis are already not required and an instability of a quantum crystal in above mentioned direction is stabilized by tubulene wall. The mentioned advantages of quantum crystals (the easy of initialization and the address to separated qubit) are saved also. One should note, that all operations with the separate qubits are carried out, as well as in case of a quantum crystal, with external constant and variable fields. To find out an ability of realization of two-qubit operations (in particular, without use of indirect interaction through an oscillatory mode) were carried out the below-mentioned quantum-chemical calculations.

3. As a model of quantum crystal the fragment of boron nitride nanotube (8,0)-type with 4 elementary layers and with introduced inside two ions $Be^+$, $Mg^+$ and atoms Li, Na and K (see fig. 1) was used. The quantum-chemical calculations of the electronic structure of the given systems were realized in the approximation of a rigid lattice by molecular cluster [8] method in the framework of semiempirical scheme MNDO [9]. The ions and atoms of metals located at the center of boron nitride nanotube fragment and its geometrical parameters optimize during computations. The calculations of single and triple configurations of atoms and ions inside tube (table 1) were carried out. The basic received results of the calculations of the geometry structure and electron-energy spectrum (distance between atoms and spectral characteristics) of boron nitride nanotubes, filled by alkaline and alkaline-earth metal atoms, are represented in table 1. The energy characteristics such as the energy of interkalicity (doping) of systems in triplet ($E_T$) and in singlet ($E_S$) states, the energy of singlet-triplet transition $\Delta E_{TS}$ are given in eV, the geometrical parameters (metal sublattice constant $R_{AB}$ in singlet (spin S=0) and in triplet (S=1) states) are expressed by angstroms.

It has appeared, that single states of the pairs of ions and atoms are energetically more advantage. The difference between the energies of singlet and triplet states, for example, for ions $Mg^+$ makes 0.11 eV, that is, on our sight, relative small value and can testify on possibility of practical using of studying system as a quantum qubit. The distance between atoms of metals practically does not depend from multiplicity of system that speaks for the benefit of mentioned application of these materials again.

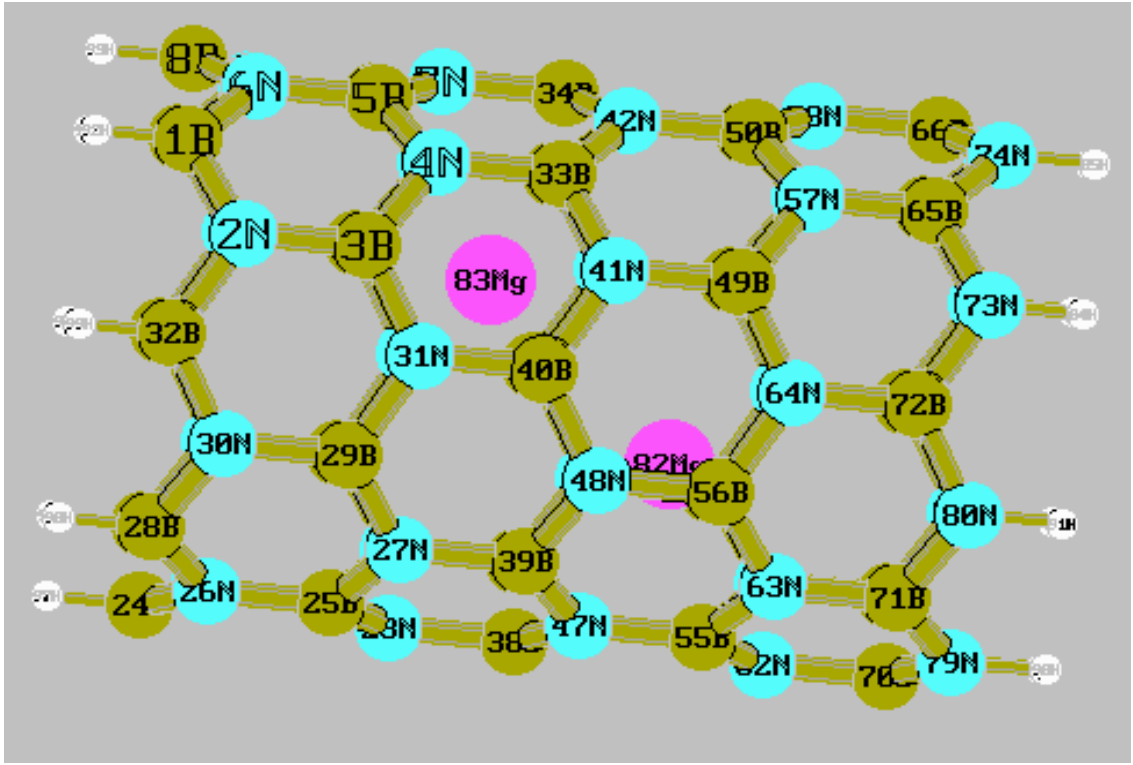

Fig. 1. The fragment of boron nitride nanotubes (8,0)-type doped by two ions Mg$^+$ with boundary atoms of hydrogen.

| Cluster | $E_T$, eV | $E_S$, eV | $R_{AB}$, Å (S=1) | $R_{AB}$, Å (S=0) | $\Delta E_{TS}$, eV |
|---|---|---|---|---|---|
| (8,0)@2Li | 4.41 | -0.91 | 2.68 | 2,10 | 0.89 |
| (8,0)@2Na | -4.79 | -21.07 | 2.40 | 2,40 | 0.24 |
| (8,0)@2K | 0.12 | 0.57 | 3.88 | 3,70 | 0.18 |
| (8,0)@2Be$^+$ | 18.03 | - | 1.98 | - | 0.01 |
| (8,0)@2Mg$^+$ | 3.30 | - | 3.07 | - | 0.11 |

Table 1. The energy and geometric characters of BN-nanotubes "zig-zag"-type, doped by pairs of alkaline atoms and ions of alkali-earth metal.

4. So in the given model there is the exchange interaction of Heisenberg type between spins of ions with characteristic energy about 0.11 eV (for ions $Mg^+$), and the basic spin state of doped nanotubes is singlet (i.e. anti-ferromagnetic type of ordering). This circumstance allows make a conclusion that the two-qubit interaction can be carried out directly without use of oscillatory modes (and, consequently, excepting the processes of the de-coherent, concerning with radiation of moving charges in oscillatory mode). The methods of realization of operation CONTROLLED NOT in two-qubit system, connected by Heizenberg interaction, are well known and for the given energy range (till 1 eV) do not represent difficulties. Besides we shall note, that the presence of anti-ferromagnetic type of the ordering allows to code the state of qubit with the help not one, but several ions [10], that raises the security from mistakes. At the same time it is necessary to note, that there is a number of difficulties, requiring additional study, in submitted approach. So, the additional mechanism of the de-coherent is possible, caused by indirect interaction of spins of atoms and ions through the phonon modes of nanotube. The oscillations of nanotube are possible to damp by means of, for example, use of a rope of nanotubes.

So the proposed model of the quantum computer on the basis of boron nitride nanotubes allows to save all advantages of the scheme with use of traps such as Paul's traps and to get rid of a number of lacks, inherent in them.

This work was supported in part by grant from RFFI (No 00-02-16445a).